\documentstyle[prl,aps,floats,epsfig]{revtex}  
\draft 

\begin{document}

\title{\large
       Measurement of Branching Fractions for $B\to \eta_c K^{(*)}$ Decays}

\maketitle



\begin{center}
  F.~Fang$^{7}$,              
  T.~Hojo$^{30}$,             
  K.~Abe$^{8}$,               
  K.~Abe$^{40}$,              
  T.~Abe$^{41}$,              
  I.~Adachi$^{8}$,            
  H.~Aihara$^{42}$,           
  M.~Akatsu$^{22}$,           
  Y.~Asano$^{47}$,            
  T.~Aso$^{46}$
% ...
\end{center}

\small
\begin{center}
% ...
$^{7}${University of Hawaii, Honolulu HI}\\
$^{8}${High Energy Accelerator Research Organization (KEK), Tsukuba}\\
% ...
$^{22}${Nagoya University, Nagoya}\\
% ...
$^{30}${Osaka University, Osaka}\\
% ...
$^{40}${Tohoku Gakuin University, Tagajo}\\
$^{41}${Tohoku University, Sendai}\\
$^{42}${University of Tokyo, Tokyo}\\
% ...
$^{46}${Toyama National College of Maritime Technology, Toyama}\\
$^{47}${University of Tsukuba, Tsukuba}\\
% ...
\end{center}

\normalsize

\normalsize

\tighten

\begin{abstract}

We report measurements of branching fractions
for charged and neutral 
$B\to \eta_c K$ decays where the $\eta_c$ meson is
reconstructed in the $K_S^0 K^{\pm}\pi^{\mp}, K^+ K^- \pi^0, K^{*0}
K^-\pi^+$ and $p \bar{p}$ decay channels. The neutral
$B^0$ channel is a $CP$ eigenstate and can be used to measure the
$CP$ violation parameter $\sin 2\phi_1$.
We also report the first observation of the
$B^0\to \eta_c K^{*0}$ mode. The results are based on an analysis of
$29.1$ fb$^{-1}$ of data collected by the Belle detector at
KEKB. 
\vskip1pc

\end{abstract}

\twocolumn[\hsize\textwidth\columnwidth\hsize\csname
@twocolumnfalse\endcsname 
\normalsize
\vskip2pc ] 

\normalsize

The decay mode $B\to \eta_c K$ proceeds by
a spectator $b\to c \bar{c} s$ transition with internal $W$-emission as in
the $CP$ eigenstate $B^0\to J/\psi K_S^0$. The neutral decay mode 
$B^0\to\eta_c K_S^0$ has therefore 
been used to measure the $CP$ violation parameter
$\sin 2\phi_1$\cite{cp_prl},\cite{cp_prd},\cite{cp_babar}. 
Measurements of branching
fractions for $B\to \eta_c K^{(*)}$ decay modes are also useful 
in the study of the dynamics of hadronic $B$ decay\cite{desh_etac}.
However, in contrast to $B^0\to J/\psi K_S^0$, the 
$\eta_c$ meson must be reconstructed from hadronic decays rather
than from a leptonic final state with relatively low combinatorial
background. In this paper, we report
new measurements of $B\to \eta_c K$ branching fractions with 
the $\eta_c$ meson reconstructed in the
$K_S^0 K^{+}\pi^{-}, K^- K^+ \pi^0, K^{*0} K^- \pi^+$ and
$p \bar{p}$ channels\cite{CC}. These signals are large enough to
be used to determine the mass and width of the $\eta_c$ meson.
We also report the first observation of the related 
decay mode $B^0\to \eta_cK^{*0}$. 
When $K^{*0}\to K^0_S \pi^0$, this decay mode is also
a $CP$ eigenstate.

We use a  $29.1~{\rm fb}^{-1}$ data sample,
which contains 31.3 million produced $B\bar{B}$ pairs, 
collected  with the Belle detector at the KEKB asymmetric-energy
$e^+e^-$ (3.5 on 8~GeV) collider~\cite{KEKB}.
KEKB operates at the $\Upsilon(4S)$ resonance 
($\sqrt{s}=10.58$~GeV) with a peak luminosity that now exceeds
$7\times 10^{33}~{\rm cm}^{-2}{\rm s}^{-1}$.
The Belle detector is a large-solid-angle magnetic
spectrometer that
consists of a three-layer silicon vertex detector (SVD),
a 50-layer central drift chamber (CDC), a mosaic of
aerogel threshold \v{C}erenkov counters (ACC), time-of-flight
scintillation counters (TOF), and an array of CsI(Tl) crystals
(ECL)  located inside 
a superconducting solenoid coil that provides a 1.5~T
magnetic field.  An iron flux-return located outside of
the coil is instrumented to identify $K_L$ and
muons (KLM).  The detector
is described in detail elsewhere~\cite{Belle_nim}.

We select well measured
charged tracks with impact parameters with respect 
to the interaction point of less than 0.5 cm in the radial direction
and less than 3 cm in the beam direction ($z$).
These tracks are required to
have $p_T>50$ MeV/$c$.

Particle identification likelihoods for the pion and kaon
particle hypotheses are calculated by combining
information from the TOF and ACC systems with $dE/dx$ measurements in
the CDC.
To identify kaons (pions), we apply a mode-dependent requirement
on the kaon (pion) likelihood ratio, $L_K/(L_\pi+L_K)$ 
($L_\pi/(L_\pi+L_K)$). The requirement
 $L_K/(L_\pi+L_K) >0.5$ is used for the $\eta_c \to K_S^0 K^- \pi^+$
 mode. For other modes, we require $L_K/(L_\pi+L_K)>0.6$, which 
 is 88\% efficient for
kaons with a $8.5\%$ misidentification rate for pions.
For the $\eta_c\to K^+ K^- \pi^0$
mode, the kaon likelihood ratio is required to be greater than
0.8 for those charged kaons that come directly from the $B$ rather than
from the $\eta_c$ candidate.
In addition, we remove all 
kaon candidates that are consistent with being either protons or electrons.

Protons and antiprotons are identified using all particle ID
systems and are required to have proton likelihood ratios 
($L_p/(L_p+L_K)$ and $L_p/(L_p+L_{\pi})$) greater than
0.4. Proton candidates that are electron-like according to the
information recorded by the CsI(Tl) calorimeter are vetoed.
This selection is $95\%$ efficient for protons
with a $12\%$ kaon misidentification rate.

We select $K_S$ candidates
from $\pi^+\pi^-$ candidates that lie within the mass window
$0.482 {\rm ~GeV}/c^2<M(\pi^+\pi^-)<0.514$ GeV/$c^2$ ($\pm 4 \sigma$). 
The flight length of the $K_S$ is required to
be greater than 0.2 cm. The difference in the
angle, in the $x-y$ plane, 
between a vector from the beam spot to the $K_S$
vertex and the $K_S$ flight direction  is
required to satisfy $\Delta\phi<0.1$ rad.

$K^{*0}$ candidates are reconstructed in the $K^+\pi^-$ mode.
For $\eta_c\to K^{*0} K^-\pi^+$, we require the $K^+ \pi^-$ 
invariant mass to be between 0.817 and 0.967 GeV/$c^2$. For the
$B^0\to \eta_c K^{*0}$ mode, the $K^{*0}$ mass must lie in
the range between 0.801 and 0.991 GeV/$c^2$.

Neutral pion candidates are selected from pairs of ECL clusters
with invariant mass within $\pm 16$ MeV of the nominal $\pi^0$ mass
and momenta above $350$ MeV/c. The photons must have energy above 50 MeV
if they lie in the barrel region of the calorimeter
and above 200 MeV if they are detected in the endcap.

To reconstruct signal candidates in the $B^+\to  \eta_c K^+$ 
and $B^0\to \eta_c K^0$ modes, 
we form combinations of charged or neutral kaons and $\eta_c$ candidates.
The $\eta_c$ is reconstructed in the 
$K_S^0 K^{+}\pi^{-}, K^- K^+ \pi^0, K^{*0} K^- \pi^+$ and
$p \bar{p}$ decay modes. The $\eta_c$ candidate
is required to have invariant mass in the range 
$2.920<M_{\eta_c}<3.035$ GeV/$c^2$ for the $K^- K^+\pi^0$ mode and
$2.935<M_{\eta_c}<3.035$ GeV/$c^2$ for all other modes.
The charged daughters of the $\eta_c$ are required to satisfy
a vertex constrained fit with a 
mode-dependent $\chi^2$ requirement\cite{chisq}.

To isolate the signal, we form the beam-energy constrained mass, 
 $M_{\rm bc}=\sqrt{E_{\rm beam}^2-\vec{P}_{\rm recon}^2}$, and
energy difference $\Delta E= E_{\rm recon}-E_{\rm beam}$
in the $\Upsilon(4S)$ center of mass frame. Here $E_{\rm beam}$,
$E_{\rm recon}$ and $\vec{P}_{\rm recon}$
are the beam energy, the reconstructed energy and the reconstructed
momentum of the signal candidate, respectively.
The signal
region for $\Delta E$ in all modes except for $\eta_c\to K_S^0 K^-\pi^+$
is $\pm 2.5\sigma$, where $\sigma$ is
the mode-dependent resolution and ranges from $\pm 25$ MeV for
$\eta_c\to p \bar{p}$ to the range $-55, +45$ MeV for
$\eta_c\to K^- K^+\pi^0$. In the low background $\eta_c\to K_S^0 K^- \pi^+$
mode, the region is extended to $\pm 35$ MeV, ($\pm 3.5 \sigma $). 
The signal region for $M_{\rm bc}$ is $5.270 {\rm ~GeV}/c^2<M_{\rm bc}<5.290$
 GeV/$c^2$. The resolution
in beam-energy constrained mass is 2.8 MeV/$c^2$ and is dominated by the
beam energy spread of KEKB.

Several event topology variables provide discrimination
between the large 
continuum ($e^+e^-\to q \bar{q}$, where $q= u, d, s, c$) 
background, which  tends to be collimated along the original quark 
direction, and more spherical $B\bar{B}$ events.
We first remove events with $R_2>0.5$, where $R_2$ is the normalized
second Fox-Wolfram moment. We form a likelihood
ratio using two variables. Six modified Fox-Wolfram
moments and the cosine of the thrust angle are combined into a Fisher 
discriminant\cite{SFW}. For signal MC and continuum data,
we then form probability density functions 
for this Fisher discriminant,
 and the cosine of
the $B$ decay angle with respect to the $z$ axis ($\cos\theta_B$).
The signal (background) 
probability density functions are multiplied together to form a signal
(background) likelihood ${\cal L}_S$ (${\cal L}_{BG})$. A mode-dependent
 likelihood ratio requirement
${\cal L}_S/({\cal L}_S+{\cal L}_{BG})$ is then imposed. 

Using a sample of 57 million $B\bar{B}$ Monte Carlo events
with a model of $b\to c$ decays, we investigate backgrounds
from other $B$ decay modes. In the $\eta_c\to p \bar{p}$ mode,
no such backgrounds are found. In other modes, some background is
observed but it can be removed 
by application of mode-dependent vetos on invariant mass combinations that
are consistent with the $D$, $D_s$, $\chi_{c1}$, $J/\psi$,
$\psi(2S)$ or $\eta_c(2S)$ masses.
For example, in the $B^+\to \eta_c K^+$, $\eta_c\to K^+ K^- \pi^0$
mode, we find there is background from the decay chain
$B^+\to \bar{D^0}\rho^+$, $\bar{D}^0\to K^- K^+$, $\rho^+\to
\pi^+\pi^0$. This background is removed by requiring that
the $K^- K^+$ invariant mass be inconsistent with the $D^0$ mass.

We fit the
$M_{\rm bc}$ distribution to the sum of a signal Gaussian and a background
function that behaves like phase space near the
 kinematic boundary\cite{argus_fun}. The width of the
Gaussian is fixed from MC simulation while the mean
is determined from $B^+\to \bar{D}^0\pi^+$ data. The shape parameter of 
the background function is determined from $\Delta E$ sideband
data.  The signal yield was determined by fits to 
the individual $M_{\rm bc}$ distributions for each mode. 
The yields
and significances\cite{significance} for these fits are given in Table~I.
Significant signals are observed in all decay modes
except for $B^0\to \eta_c K^0$, $\eta_c \to K^{*0} K^-\pi^+$.
For this mode, we calculate an 
upper limit based on the number of events observed
in the $M_{\rm bc}$ signal region (4) and the expected number of
background events (2)  based on the fit. We use the
Feldman-Cousins procedure\cite{feldman}, and reduce the efficiency
by one sigma of the systematic error in the calculation\cite{cousins}.
The detection efficiencies for all modes were determined from a 
GEANT based Monte Carlo simulation.

\begin{figure}[htb]
\centerline{
\epsfxsize 1.6 truein \epsfbox{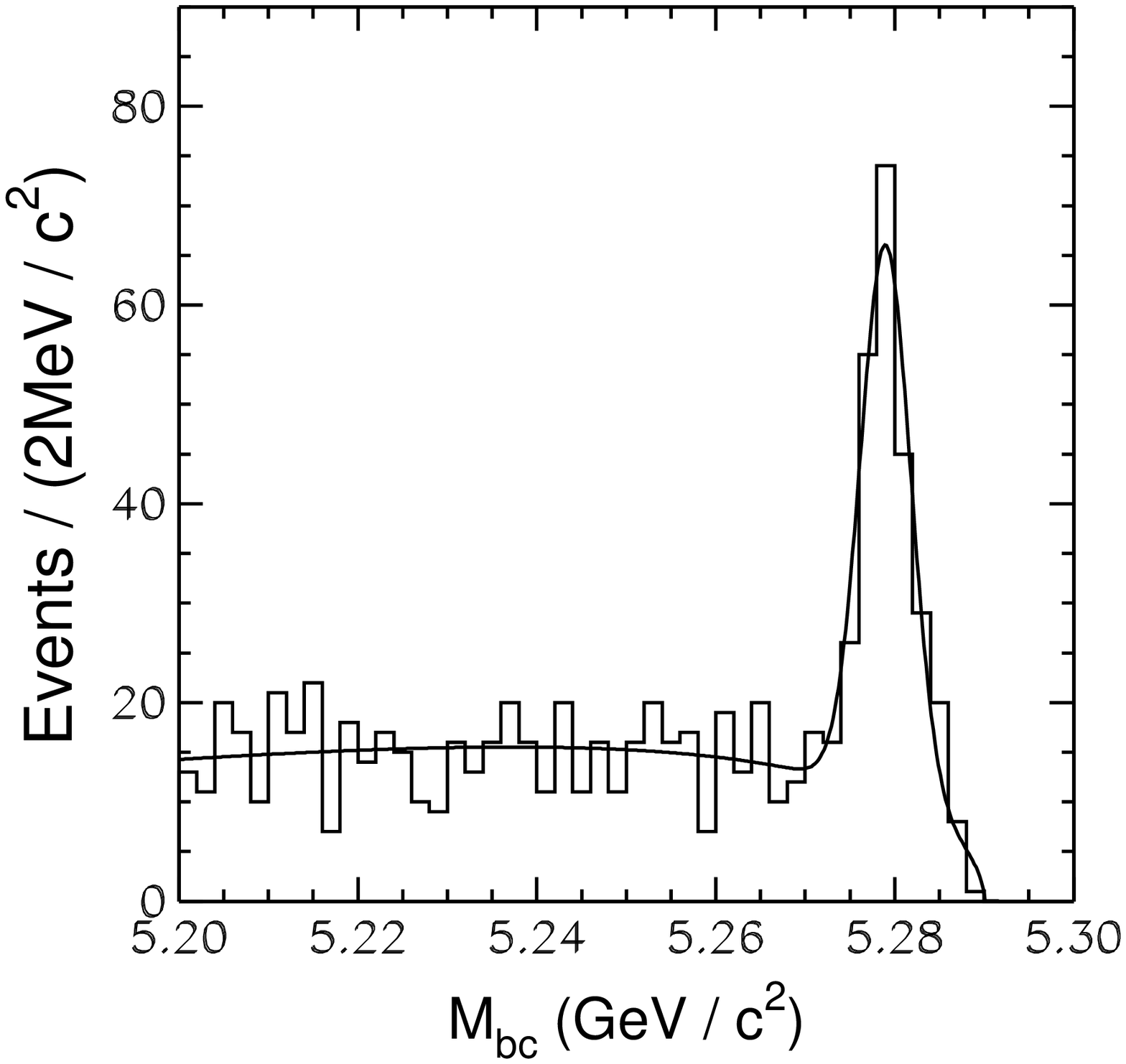}
\epsfxsize 1.6 truein \epsfbox{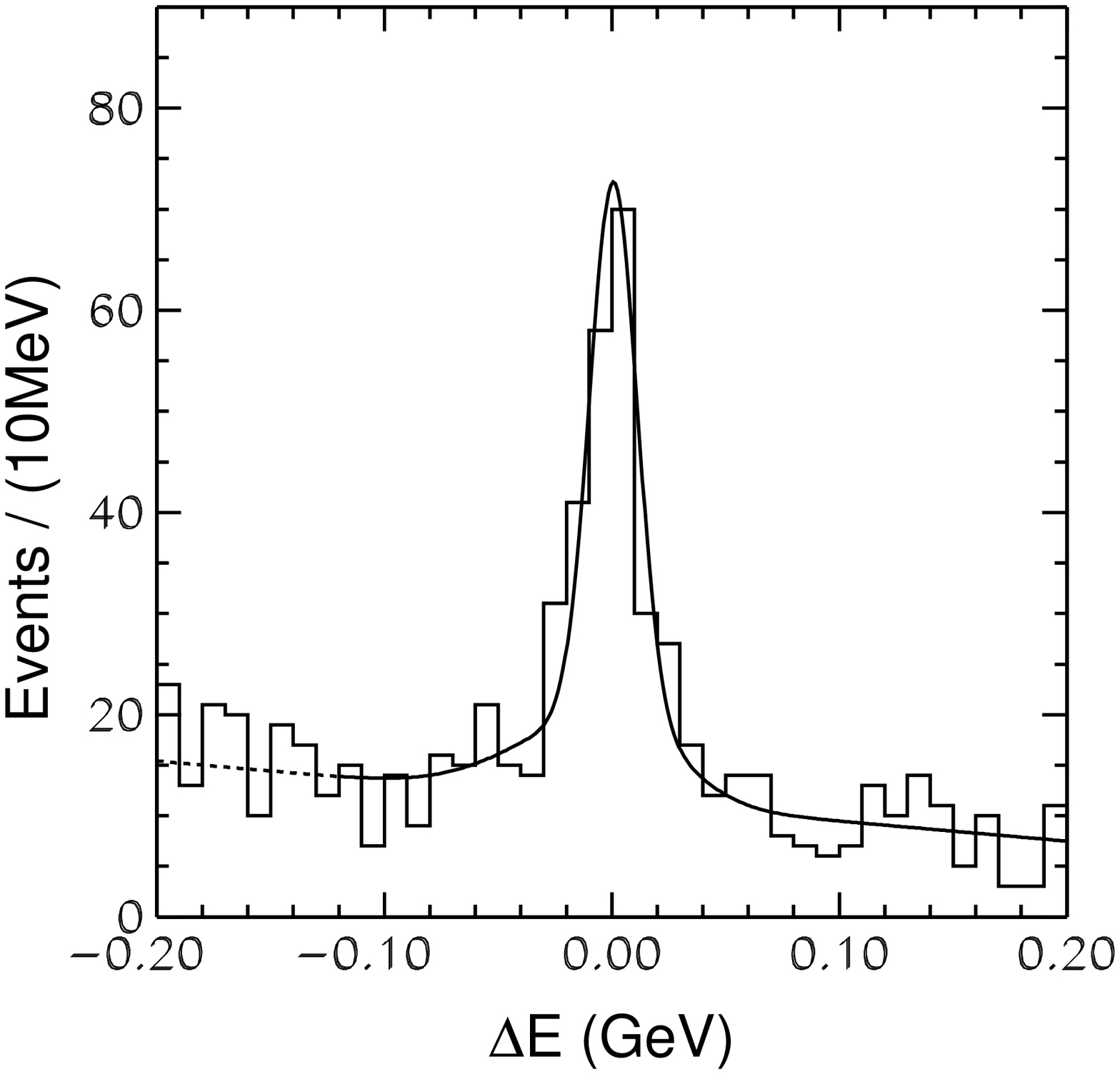}
}   
\caption{(a) $M_{\rm bc}$ and (b) $\Delta E$
distributions for         
$B\to \eta_c K$ candidates in all the decay modes.}
\label{mbde_etack}
\end{figure}

For illustration, in Fig.~\ref{mbde_etack}, we show the 
beam-energy constrained mass and $\Delta E$ 
distributions for the signal candidates in
all the decay modes except for $B^0\to \eta_c K_S^0$, 
$\eta_c\to K^{*0} K^-\pi^+$. 
In the $M_{\rm bc}$ distribution, 
we observe a signal of $195\pm 17$ events.

As a consistency
check, we also determine the yield from a fit to the $\Delta E$
distribution with a double Gaussian for signal 
and a linear background function with slope determined from
the $M_{\rm bc}$ sideband. The results of these fits are also given 
in Table I.
The fit to the $\Delta E$ distribution for all modes
combined gives an integrated yield
of $188\pm 17$ events in the signal region.
 In the fits to the $\Delta E$ distribution,
the region with $\Delta E<-120$ MeV is excluded to avoid contributions
from modes with additional particles such as $B \to \eta_c K^{*}$. 

\begin{figure}[htb]
\centerline{\epsfxsize 1.9 truein \epsfbox{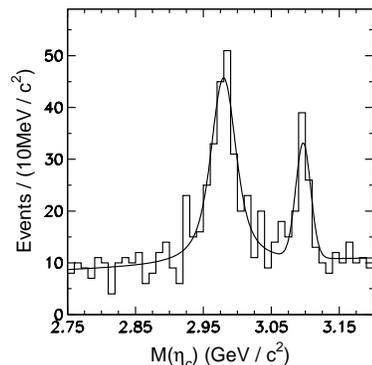}}   
\caption{Candidate $M(\eta_c)$ invariant mass distribution
for events in the $M_{\rm bc}$ and $\Delta E$       
signal region. Signals at the $\eta_c$ and $J/\psi$ masses 
from $B\to \eta_c K$ and $B\to J/\psi K$ decays are visible.}
\label{metac_etack}
\end{figure}

After removing the requirements on $\eta_c$ invariant mass,
we also verify that the signal yield for $\eta_c$ candidates
in the $M_{\rm bc}$, $\Delta E$ signal region is consistent with
the result used for the branching fraction determination.
The $\eta_c$ invariant mass distribution for signal candidates
is shown in Fig.~\ref{metac_etack}. Fitting to a Breit-Wigner
convolved with the resolution determined from MC, we find 
an intrinsic
width $\Gamma(\eta_c)=29\pm 8\pm 6$ MeV and a mass $M(\eta_c)=2979.6\pm
2.3\pm 1.6$ MeV. The systematic errors in the width and mass
measurements
include the effects of varying the background shape, the small difference
between data and MC detector resolutions, and possible binning effects.
The results are consistent with world averages and comparable in precision
to the best individual measurements\cite{PDG}. 
The yield is $182\pm 25$ events. 
We also observe a clear
signal of $66\pm 18$ events
from $B\to J/\psi K$, where the $J/\psi$ is reconstructed
in hadronic decay modes, that has a $J/\psi$ mass
and yield  that are consistent with  values obtained
for the $J/\psi \to \ell^+\ell^-$ decay mode.

The contributions to the systematic error include
the uncertainties due to the tracking
efficiency (2\% per track), particle identification efficiency
(4-14\%, depending on the mode) 
and the modeling of the likelihood ratio requirement (2\%).
The error in kaon identification 
efficiency is obtained from kinematically selected 
$D^{*+}\to D^0\pi^+, D^0\to K^-\pi^+$ in the data while 
the error in proton/antiproton
identification is determined using
$\Lambda/\bar{\Lambda}$ samples. The systematic
error due to the modeling of the likelihood ratio cut
is determined using $B^+\to \bar{D}^0\pi^+$ events reconstructed in data.
The systematic error in the yields of the $M_{\rm bc}$ fit  
were determined by varying the mean and $\sigma$ of the signal and the shape 
parameters of the background. To account for the possibility of background
from non-resonant modes that may contribute to the $M_{\rm bc}$
distribution, we include the yields observed in the
$\eta_c$ mass sideband (8-14\% of the signal depending on the mode) 
as an asymmetric systematic error.
The sources of systematic error
are combined in quadrature to obtain the total systematic error,
which is given in Table II.

\begin{table}[htb]
\label{yields}
\begin{center}
\caption {Signal yields from $M_{\rm bc}$ and
$\Delta E$ fits, statistical significances
 and MC reconstruction efficiencies. Errors are statistical only.}
\begin{tabular} {l|c|c|c|c} 
${B}^{+} \rightarrow \eta_{c}K^{+}$ & $\Delta{E}$ Yield &
$M_{\rm bc}$ Yield & Signif.($M_{\rm bc}$) & $\epsilon (\%)$ \\
$\eta_{c} \rightarrow K_S^0 K^{-}\pi^{+}$ & $74.8 \pm 10.4$  & $81.6\pm
10.3$  & 12.2$\sigma$ & $16.4$ \\
$\eta_{c} \rightarrow K^{+}K^{-}\pi^{0}$ & $26.5\pm 7.8$ &
$31.8\pm7.0$ & 6.3$\sigma$ & $8.8$ \\
$\eta_{c} \rightarrow p\bar{p}$ & $16.3\pm4.2$ & $17.7\pm4.4$ &
7.5$\sigma$ &$34.0$ \\
$\eta_{c} \rightarrow K^{* 0}K^{-}\pi^{+}$ & $22.0\pm5.8$ &
$20.8\pm5.4$ & 5.8$\sigma$ & $5.1$ 
\end{tabular}

\begin{tabular} {l|c|c|c|c} 
${B}^{0} \rightarrow \eta_{c}K_S^0 $ & $\Delta{E}$ Yield &
$M_{\rm bc}$ Yield &
Signif.($M_{\rm bc}$) & $\epsilon (\%)$ \\
$\eta_{c} \rightarrow K_S^0 K^{-}\pi^{+}$ & $19.6\pm 5.4$  & $23.0\pm
5.4$ & 6.8$\sigma$ & $15.5$ \\
$\eta_{c} \rightarrow K^{+}K^{-}\pi^{0}$ & $19.9\pm5.7$
&$17.1\pm5.1$& 4.7$\sigma$ & $9.5$ \\
$\eta_{c} \rightarrow p\bar{p}$ & $7.0\pm3.0$ & $6.8\pm2.6$ &
5.0$\sigma$ & $34.9$  \\
$\eta_{c} \rightarrow K^{* 0}K^{-}\pi^{+}$ & $0.2\pm 1.7$ &
$2.2\pm 1.8$ & 1.6$\sigma$ & $3.75$ 
\end{tabular}

\end{center}
\end{table}

\begin{table}[htb]
\label{productbr}
\begin{center}
\caption {Product branching fractions for $B\to \eta_c K$
decay modes ($10^{-6}$).}
\begin{tabular}{l l} 
$\mathcal{B}$$(B^{+} \rightarrow \eta_{c}K^{+}) \times
\mathcal{B}$$(\eta_{c} \rightarrow K_S^0 K^- \pi^{+})$ &
$(23.2\pm 2.9^{+2.8}_{-3.8})$ \\
$\mathcal{B}$$(B^{+} \rightarrow \eta_{c}K^{+}) \times
\mathcal{B}$$(\eta_{c} \rightarrow K^{+}K^{-}\pi^{0} )$ &
$(11.4\pm 2.5^{+1.1}_{-1.8})$ \\
$\mathcal{B}$$(B^{+} \rightarrow \eta_{c}K^{+}) \times
\mathcal{B}$$(\eta_{c} \rightarrow p\bar{p})$ &
$(1.64\pm 0.41^{+0.17}_{-0.24})$ \\
$\mathcal{B}$$(B^{+} \rightarrow \eta_{c}K^{+}) \times
\mathcal{B}$$(\eta_{c} \rightarrow K^{* 0}K^{-}\pi^{+})$ &
$(19.3\pm 5.0^{+3.4}_{-3.8})$ \\
&\\
$\mathcal{B}$$(B^{0} \rightarrow \eta_{c}K^{0}) \times
\mathcal{B}$$(\eta_{c} \rightarrow K_S^0 K^- \pi^{+})$ &
$(20.1\pm 4.7^{+3.0}_{-4.5})$ \\
$\mathcal{B}$$({B}^{0} \rightarrow \eta_{c}K^{0}) \times
\mathcal{B}$$(\eta_{c} \rightarrow K^{+}K^{-}\pi^{0})$ &
$(16.6\pm 5.0^{+1.8}_{-1.8})$ \\
$\mathcal{B}$$({B}^{0} \rightarrow \eta_{c}K^{0}) \times
\mathcal{B}$$(\eta_{c} \rightarrow p\bar{p})$ & 
$(1.79\pm 0.68^{+0.19}_{-0.25})$ \\
$\mathcal{B}$$({B}^{0} \rightarrow \eta_{c}K^{0}) \times
\mathcal{B}$$(\eta_{c} \rightarrow
K^{*0}K^{-}\pi^{+})$ & $(8.1\pm 6.6\pm 1.4) $ \\
    & $<29$ at 90\% C.L.

\end{tabular}
\end{center}
\end{table}


The product branching fractions are given in Table II for all modes in which
signals are observed. Since many of the $\eta_c$ branching fractions
are poorly determined and in some cases there are conflicting measurements,
 we quote $B$ branching fractions for 
the $\eta_c\to K_S^0 K^-\pi^+$ and $\eta_c\to K^- K^+ \pi^0$ 
modes only. The $\eta_c\to K_S^0 K^-\pi^+$  mode is the most
precisely and reliably measured mode and the 
branching fraction for the $\eta_c\to K^- K^+\pi^0$ 
mode is related by isospin. 
We use ${\cal B}(\eta_c\to K_S^0 K^-\pi^+)=1/3 \times
(0.055\pm 0.017)$ where $1/3$ is the product of the appropriate Clebsch-Gordon
coefficient and intermediate $K^0$ branching fraction. 
We assume that the experimental systematic errors in
the $\eta_c\to K_S^0 K^- \pi^+$ and $\eta_c\to K^- K^+ \pi^0$ modes
are uncorrelated. We assume equal production of $B^+B^-$ 
and $B^0\bar{B}^0$ pairs and do not include an additional systematic
error for the uncertainty in this assumption.
%
%
%
We find $${\cal B}(B^+\to \eta_c K^+)= (1.25\pm 0.14^{+0.10}_{-0.12}\pm
0.38)\times 10^{-3}$$ and
$${\cal B}(B^0\to \eta_c K^0)= (1.23\pm 0.23^{+0.12}_{-0.16}\pm 0.38)
\times 10^{-3}.$$

The first error is statistical, the second error is systematic
and the third error is due to the uncertainty in the $\eta_c$
branching fraction scale.
When the $\eta_c$ branching fractions for the other modes
are better determined, 
absolute $B$ branching fractions for these modes can
be extracted from our results.

In the $B^0\to \eta_c K^{*0}$, $K^{*0}\to K^+\pi^-$ 
channel, the $\eta_c$ is reconstructed in
the $K_S^0 K^{\pm}\pi^{\mp}$ mode. Since this mode is a pseudoscalar to 
pseudoscalar-vector decay, by angular momentum conservation,
the cosine of the $K^*$ helicity angle ($\cos\theta_H$) follows a 
$\cos^2\theta_H$ distribution; we select events with
$|\cos\theta_H|>0.4$.
We also investigate $B\bar{B}$ background and find that this
background can be removed by applying vetos to events with combinations
that are consistent
with $J/\psi\to K_S^0 K \pi$, $J/\psi \to K^+ K^- \pi^+ \pi^-$,
$\psi(2S)\to K_S^0 K \pi$, $\eta_c(2S)\to K_S^0 K \pi$,
 $\chi_{c1}\to K_S^0 K \pi$, $\eta_c\to K^+ K^- \pi^+ \pi^-$, 
 $D_s^+\to K_S^0 K^+$ and 
$D_s\to K^- K^+ \pi$. The detection efficiency for these
selection requirements is $(7.95\pm 0.12) \%$.

\begin{figure}[htb]
\centerline{
\epsfxsize 1.8 truein \epsfbox{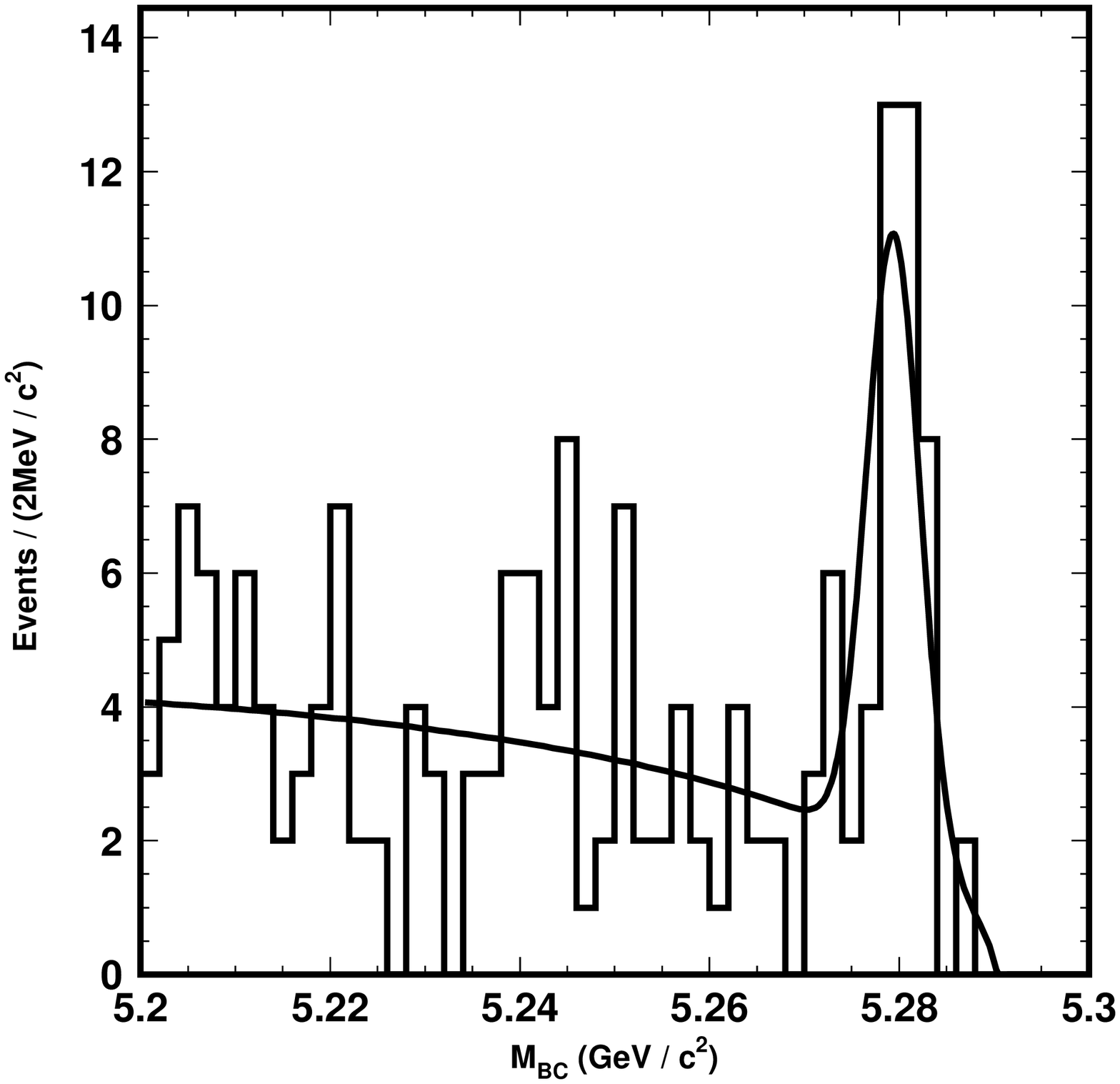}
\epsfxsize 1.8 truein \epsfbox{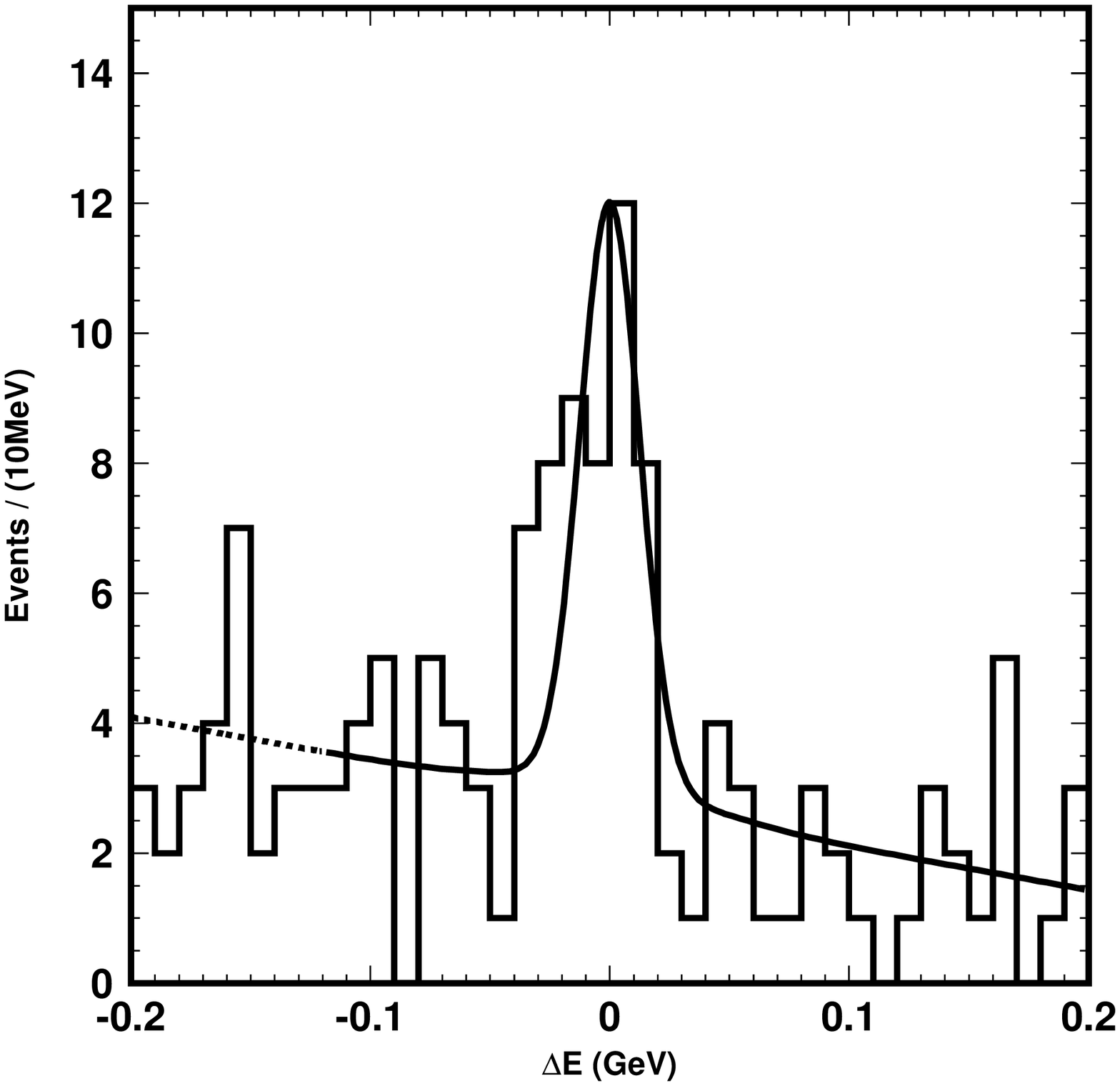}
}   
\caption{(a) $M_{\rm bc}$  and (b) $\Delta E$ distributions
for $B\to \eta_c K^{*0}$ candidates.}
\label{etackst}
\end{figure}

After applying these requirements,
a fit to the $M_{\rm bc}$ distribution yields
a signal of $33.7\pm 6.7$ events for $B^0\to \eta_c K^{*0}$
with a statistical significance of $7.7\sigma$.\cite{significance} 
The $M_{\rm bc}$ distribution is shown in Fig.~\ref{etackst}(a).
The yields from the
$\Delta E$ fit, shown in Fig. 3(b) ($30\pm 7$ events), 
%
the $\eta_c$ invariant mass distribution ($24\pm 7$ events) and
the $K^+\pi^-$ invariant mass distribution
($27\pm 8$ events) are consistent with the yield from the
$M_{\rm bc}$ fit. 

To evaluate the contribution from non-resonant 
$B^0 \to K_S^0 K^{+} \pi^{-} K^{*}$  as well as
the remaining $B\overline{B}$ backgrounds that peak in
the $M_{\rm bc}$ distribution, we select events in
the $\eta_c$ sideband\cite{sideband} and repeat the $M_{\rm bc}$ fit.
We find no significant signal. By using the ratio
of the yields in the $\eta_c$ signal and sideband regions
determined from MC, we estimate the contributions of such
backgrounds 
to be $3.9\pm 4.2$ events, consistent with zero.
We use the $K^*$ sideband\cite{sideband} to estimate
the non-resonant $B^0 \to \eta_c K\pi$ decay component and
obtain $-0.6\pm 3.3$ events.
These possible background contributions are not subtracted in the branching
fraction measurement, but instead are treated as systematic uncertainties.
We find 
$${\cal B}(B^0\to \eta_c K^{*0}) = (1.62\pm 0.32^{+0.24}_{-0.34}\pm
0.50)\times 10^{-3}.$$
To  take into account the possibility of $B\bar{B}$
background, we conservatively include asymmetric systematic errors from
the results of the fits to the $\eta_c$($-12\%$) and $K^*$ ($-8\%$) 
sidebands. Other sources of systematic error are the uncertainties
in the track reconstruction efficiency ($\pm 2\%$ per track),
the parameters in the $M_{\rm bc}$ fit ($\pm$ 7\%), particle
identification ($\pm 6\%$) and the number of $B^0\bar{B}^0$ events.


From the results for the branching fractions for
$B^0\to \eta_c K^0$ and $B^0\to \eta_c K^{*0}$ determined above,
we can determine the ratio 
$ R_{\eta_c} = {\cal B}(B^0\to \eta_c K^{*0})/{\cal B}(B^0\to \eta_c
K^0)$. The uncertainty from the $\eta_c$ branching fraction scale
cancels in the ratio.
We find $$R_{\eta_c} = 1.33\pm 0.36^{+0.24}_{-0.33}.$$
Our result can be compared to calculations of this ratio in models
based on factorization and is consistent with the range 1.02-2.57
predicted by Gourdin, Keum and Pham.\cite{theory_R}.
%
%

The branching fractions reported in this paper for $B^+\to\eta_c K^+$
and $B^0\to \eta_c K^0$ using $\eta_c\to K \bar{K} \pi$ decays
are more precise than previous results\cite{CLEO_etack}. The result
for the $B^+\to\eta_c K^+$ branching fraction 
is somewhat higher than the CLEO measurement, while the
$B^0\to \eta_c K^0$ result is consistent.
Several additional $\eta_c$ modes
including $\eta_c\to p \bar{p}$, $\eta_c\to K^- K^+\pi^0$ and 
$\eta_c\to K^{*0} K^-\pi^+$ have
been used and increase the fraction of $\eta_c$ decays that can
be reconstructed for $CP$ violation
measurements. With the large samples of $B\to \eta_c K$ decays now
available, we are able to determine the mass and width
of the $\eta_c$ meson. We find $M(\eta_c)=2979.6\pm
2.3\pm 1.6$ MeV and $\Gamma(\eta_c)=29\pm 8\pm 6$ MeV.
In addition, we report the first observation of
the $B^0\to \eta_c K^{*0}$ decay, which is an $CP$ eigenstate
when $K^{*0}\to K_S^0\pi^0$.


We wish to thank the KEKB accelerator group for the excellent
operation of the KEKB accelerator.
We acknowledge support from the Ministry of Education,
Culture, Sports, Science, and Technology of Japan
and the Japan Society for the Promotion of Science;
the Australian Research Council
and the Australian Department of Industry, Science and Resources;
the National Science Foundation of China under contract No.~10175071;
the Department of Science and Technology of India;
the BK21 program of the Ministry of Education of Korea
and the CHEP SRC program of the Korea Science and Engineering Foundation;
the Polish State Committee for Scientific Research
under contract No.~2P03B 17017;
the Ministry of Science and Technology of the Russian Federation;
the Ministry of Education, Science and Sport of Slovenia;
the National Science Council and the Ministry of Education of Taiwan;
and the U.S.\ Department of Energy.


\begin{thebibliography}{99}
\bibitem{CC} The inclusion 
of the charge conjugate mode is implied.
\bibitem{chisq} The vertex $\chi^2$ requirement is 
$<50$ for all modes except for $\eta_c\to K_S^0 K^- \pi^+$,
where a less restrictive cut $<100$ is used.
\bibitem{significance} The significances are calculated from
the change in the likelihood when the signal yield is set to zero.
\bibitem{sideband} The mass regions 
$2.74<M_{K_S K^{\pm}\pi^{\mp}}<2.84$ GeV/c$^2$
and $3.25<M_{K_S K^{\pm} \pi^{\mp} }<3.35$ GeV/$c^2$ are used as the
$\eta_c$ sideband. The range $1.60<M_{K^- \pi^+}<1.79$ GeV/c$^2$ is used
as the $K^*$ sideband.
\end{thebibliography}
\end{document}